\documentclass[preprint,12pt]{elsarticle}
\usepackage{comment}
\usepackage{rotating}
\usepackage{graphicx} 
\usepackage{subfig} 
\usepackage{enumerate} 
\usepackage{amssymb}
\usepackage{geometry,mathtools}
\usepackage{latexsym}
\usepackage{dcolumn}
\usepackage{ulem}
\usepackage{multirow}
\usepackage{hyperref}
\usepackage{diagbox}
\usepackage{booktabs}
\usepackage{esvect}
\usepackage{xcolor}
\usepackage[utf8]{inputenc}
\usepackage{caption}
\usepackage{subcaption}
\usepackage{colortbl}
\usepackage[table]{xcolor}
\definecolor{lightgray}{gray}{0.9}
\usepackage{float}
\usepackage{mathrsfs}

\newcommand{\R}{$\mathfrak{R}$}
\newcommand{\RinFla}{\mathfrak{R}}
\newcommand{\Rc}{$\mathfrak{R}_0$}

\begin{document}

\begin{frontmatter}

\title{Epidemic Spread: Limiting Contacts to Regular Circles Is Not Necessarily the Safest Option.}

\author{Jo\~ao Gabriel Sim\~oes Delboni$^{a}$}
\author{Gabriel Fabricius$^{a,b}$}
\address{$^a$ Instituto de Investigaciones  Fisicoqu\'{\i}micas Te\'oricas y Aplicadas (INIFTA), CONICET and	Facultad de Ciencias Exactas, Universidad Nacional de La Plata,	CC 16, Suc. 4, 1900 La Plata, Argentina }
\address{$^b$ CCT CONICET La Plata, Consejo Nacional de Investigaciones Científicas y T\'ecnicas, Argentina}

 \begin{abstract}
When a new infectious disease (or a new strain of an existing one) emerges, as in the  
recent COVID-19 pandemic, different types of mobility restrictions are considered to
slow down or mitigate the spread of the disease. The measures to be adopted require
carefully weighing the social cost against their impact on disease control. In this
work, we analyze, in a context of mobility restrictions, the role of frequent versus occasional
contacts in epidemic spread. We develop an individual-based mathematical
model where frequent contacts among individuals (at home, work, schools) and occasional
contacts (at stores, transport, etc.) are considered. 
We define several contact structures by varying the relative weight of frequent and occasional contacts while keeping the same initial growth rate of the epidemic.
We find the remarkable result that the more frequent contacts prevail over occasional ones, the higher the epidemic peak, the sooner it occurs, and the greater the final number of individuals affected
by the epidemic.  We conduct our study using an SIR model, considering both exponential and deterministic recovery from infection, and obtain that this effect is more pronounced under deterministic recovery. We find that the impact of relaxation measures depends on the relative importance of frequent and occasional contacts within the considered social structures.
Finally, we assess in which of the considered scenarios the homogeneous mixing approximation provides a reasonable description of the epidemic dynamics.

\end{abstract}

\end{frontmatter}

\section{Introduction}

In the transmission of infectious diseases that spread through direct contact, such as measles, pertussis, and COVID-19, various types of contacts are involved. For example, an individual typically has frequent contacts with the same person (for instance, at home, with coworkers, etc.), or occasional contacts with people they seldom see or may never see again (for example, in a store, on public transportation, etc.). The relative weight of the different types of contacts that each individual has and how these are articulated in a contact network with other individuals define the contact structure of a given society through which a disease spreads.

In this work, we aim to explore the implications of certain characteristics of the contact structure for the transmission process of a given disease. Specifically, we are interested in characterizing the consequences of considering contact structures where either frequent contacts prevail over occasional ones or vice versa. This is relevant because within the same society, the relative weight between frequent and occasional contacts can sometimes be regulated or modified through control measures. For example, during the recent COVID-19 pandemic, the tension between restricting outings and the need to go out (e.g., for work) led to a variety of situations where it was not immediately obvious which entailed greater risk.

We have developed an individual-based model (IBM) and
constructed an artificial city
where we could design structures that explicitly consider frequent contacts such as in households, student groups in schools, among coworkers, or occasional contacts where individuals do not necessarily contact the same people every day, such as those in stores or transportation. The generated structures are not intended to have the level of detail of a realistic population \cite{ferguson2005}, nor to be too simplified as in some works where the goal is to obtain a theoretical approximation \cite{ball2008, dta2016}. The generated structures take into account the basic contacts that occur in the most characteristic environments without including an excessive number of parameters that could hinder the interpretation of the results and the visualization of the most important effects.

In particular, we address the following question: for an epidemic in its early stage, growing exponentially at a given rate, how does its subsequent evolution depend on the relative weight of frequent versus occasional contacts? More precisely, consider two scenarios in which the epidemic grows at the same initial exponential rate, but where a larger fraction of transmissions takes place in local (frequent-contact) settings in one case than in the other. Which of these scenarios should we expect to lead to a larger epidemic?

To answer this question, we perform a controlled comparison across scenarios with different weights of frequent and occasional contacts, ensuring that the epidemic initially grows at the same rate in all cases. In this way, differences in the observed dynamics can be attributed specifically to the underlying contact structure. Our findings reveal a nontrivial and, to some extent, counterintuitive result: epidemics tend to have a larger overall impact as the proportion of frequent contacts increases. This behavior is robust across the different structures considered and, to the best of our knowledge, has not been previously reported.

From an epidemiological perspective, this result may have relevant implications. It is well documented that social behavior and epidemic dynamics are coupled through feedback mechanisms \cite{pan2020, he2013, poletti2011}: the implementation of social distancing measures typically reduces incidence, which in turn may lead to a relaxation of precautions as the perceived risk declines, potentially resulting in renewed increases in transmission or the emergence of subsequent waves.
In this context, consider a situation in which an epidemic is spreading at a given rate and interventions are introduced to reduce occasional contacts (for instance, in public transportation or commercial settings). Our results suggest that particular care is needed to ensure that any improvement in the epidemiological situation does not lead to a relaxation of precautions in settings dominated by frequent contacts. Indeed, if such a relaxation were to occur and the epidemic were to resume spreading at a rate comparable to the original one, the overall impact of the epidemic could be worse than before the intervention.

We are also interested in addressing a methodological concern. Homogeneous mixing (HM) models have been widely used to study various aspects of infectious diseases worldwide \cite{AMlibro}. For instance, during the recent COVID-19 pandemic, they were extensively employed to estimate the impact of different control measures. However, it is well known that if local interactions are important, the HM description can lead to misleading results. 
Therefore, we aim to determine whether the homogeneous mixing approach yields reasonable predictions for any of the considered scenarios.

\section{Materials and Methods}
\label{sec2}

\subsection{The model}

We consider a stochastic individual-based SIR epidemic model where each susceptible (S) individual can become infected (I) through contact with an infected individual in a variety of locations. Recovery from infection for each individual occurs after a period of time drawn from a given distribution. Once recovered (R), individuals remain in this state until the end of the epidemic.
Note that recovered individuals have been infected in the past; however, we use ``infected'' to denote individuals who are currently infected, which in the SIR model is equivalent to being infectious.

Our aim is to model interactions among people in a city, corresponding to social dynamics with restrictions. We have set the population size at a constant N = 180,000, excluding considerations of demographic dynamics. We assume an initial number of  $I_{ini}$ individuals are infected, disregarding how they acquired the infection. From that point on, we neglect the influx and outflux of individuals into and out of the city, so we consider the dynamic evolution of the epidemic in an isolated system.

\subsubsection{Social structure and infections}
\label{socialstruct}
Each individual is assigned a set of $places$ they go to, initially determined based on their age. The age distribution is maintained uniformly constant up to 75 years and is initially used to form households ($H$) that can accommodate 1, 2, or 4 individuals. In addition to their homes, individuals aged 4 to 18 are assigned to an educational institution (school: $Sch$), where they can interact with others within the classroom ($Sch_1$) or within a smaller group of friends from the class ($Sch_2$). Those aged 20 to 65 are assigned to a workplace ($W$), and those aged 12 to 75 to a variable number (between 1 and 12) of retail stores or boutiques ($B$). The city is divided into 4 quarters ($Q$) that contain a set of households. 
An individual can, in principle, interact with any other individual within their quarter. Here we consider contacts that may occur in public spaces such as streets, parks, leisure venues, large stores, public transportation, and transport stations.
 In addition to the $places$, each individual in a household is assigned to an extended family ($F$) that consists of members from two other households.

A detailed explanation of how places and extended family are assigned to each individual is provided in the Supplementary Material (SM). To model interactions between individuals, we assume that in each place an individual is connected to all others, meaning that the individuals form a complete network in that place. Extended families have a different structure, where the members of the extended family of individual $j$, $F_j$, have other extended families assigned to them, which are different from $F_j$ (see SM). We use the term ``location'' to include both situations that give rise to contacts: places and extended family. Table \ref{enes} summarizes the different locations considered where contacts may occur.

We define an infectious contact as an interaction where, if one individual is susceptible and the other is infected, the susceptible individual becomes infected. We model infectious contact between individuals at location $l$ as a homogeneous Poisson process with a transmission rate $\beta_l$. This means that the probability of disease transmission from an infective to a susceptible individual in a time interval of
length $\Delta t$ is given by


\begin{equation} \label{pl}
p_l = 1 - e^{-\beta_l \Delta t}.
\end{equation} 

Then the probability that the susceptible individual $j$ gets infected in $\Delta t$ is:

\begin{equation} \label{Pi}
P_j = 1 - \prod_l (1-p_l)^{I_{l}}= 1-e^{-\lambda_j \Delta t} ~~{\rm with}~\lambda_j=\sum_l \beta_l I_{l},
\end{equation}

\noindent where $l$ in the product runs over the locations assigned to individual $j$ and $I_l$ is the total number of infected individuals at the respective location \cite{bansal2007}. We have chosen  $\Delta t$ = 1 day, so $\beta_l$ represents an effective average transmission rate, considering that individuals spend certain times of the day at different locations (for example at work or home). We do not consider heterogeneities arising from locations that are more frequented on weekends (such as leisure places included in $Q$ or those in $F$).

For the transmission rate at location $l$, we assume the following dependence on 
the number of individuals that could be contacted, $N_l$ (for $N_l>0$):
\begin{equation}
\label{betal}
\beta_l= \left\{
	   \begin{array}{ll}
	   \beta_T             & {\rm when} ~ l~ {\rm is~ type~}  T=H, F\\
	   \frac{\beta_T}{N_l} & {\rm when} ~ l~ {\rm is~ type~} T=W, Sch, B, Q. \\
	   \end{array} \right.
\end{equation}

That is, we assume that within households, individuals interact with all other members with the same rate regardless of the household size,
i.e., the number of contacts  per individual increases with household size (density-dependent transmission). 
In contrast, in workplaces, as the workgroup size increases, the per-individual contact rate decreases, while the total rate of contacts remains the same (frequency-dependent transmission) 
\cite{VanBoven2007}.
In Eq.(\ref{betal}) $T$ denotes the type of location that may be $H,F, W, Sch, B$ or $Q$.
It is worth noting that, in the case of retail stores or boutiques ($B$), Eqs.(\ref{pl}-\ref{betal}) are only valid under the assumption that individuals frequent a single store. However, as explained in the SM, each individual can visit a different number of stores. On a given day, an infected individual chooses one of these stores, where they can  infect other individuals. Therefore, the interaction at a given store, $l$, between two individuals $i$ and $j$ becomes dependent on the specific pair ($\beta_l \rightarrow \beta_l^{(ij)}$). While the general expressions are provided in the SM, we chose to present these simplified forms for the sake of clarity in the main text, under  the particular assumption that each of the two individuals has only one $B$-type location to choose from.

This approach of maintaining the total transmission rate of the same order for all individuals, including those with many contacts, was adopted to avoid introducing pronounced heterogeneity in contact intensity across individuals. Although such effects have been suggested to be important in some contexts \cite{britton_heter,MGMG}, we deliberately chose not to include them, as they could obscure our main objective of assessing the role of frequent versus occasional contacts.

Finally, it should be highlighted that age is used solely to build the contact structure, as it determines the activities individuals engage in and the places they go to.
While for many diseases infectivity and susceptibility may vary with age, such dependences are not accounted for in this model.

\subsubsection{Recovery from infection}

The time $\tau_j$ individual $j$ remains infected is taken randomly from a given distribution.
In this work, we consider two cases: either the $\tau_j$ are chosen with probability $p(\tau_j) \propto \exp(-\tau_j/\tau_{inf})$ (exponential recovery, ER), or they are all equal to $\tau_{inf}$, i.e., each individual recovers after a fixed time (deterministic recovery, DR).
We considered these two types of recovery because they often represent two limiting cases, with most infectious diseases exhibiting an intermediate behavior \cite{KRlibro}. Furthermore, it has been highlighted that certain properties of the dynamics can be strongly dependent on the type of recovery \cite{lloyd2001}.

\subsection{Simulations}

At the beginning of the simulation ($t=0$), all individuals are assigned the state S except for a number $I_{ini}$ of individuals chosen uniformly at random, who are assigned the state I.
We also assign to each individual $j$, a time $\tau_j$
(taken from the corresponding distribution of recovery times) that
will be used in case this individual becomes infected. We simulate the system dynamics with a discrete time step, $\Delta t$, of a day. At each day, $t \geq 0$, the state of the system is defined by the epidemiological state, $X_j $= S, I, or R, of each individual $j$, in the city, and the time $t_k$ that has passed since each infected individual $k$ has acquired infection.
We denote $S(t)$, $I(t)$, $R(t)$ the total number of individuals in the system 
that are in the states S, I, and R at time $t$, respectively, and 
$s(t)$, $i(t)$, $r(t)$ the corresponding fractions that are in these states.
The new infections produced in the system, $n_{inf}(t)$, by the $I(t)$
infected individuals present at the beginning of the day $t$ are obtained following the procedure described in the SM. Then, the time since getting infected for each infectious individual, $t_k$, is increased by $\Delta t$, and the number of
infected individuals recovered this day, $n_{rec}(t)$, is computed as the number
of infected individuals $k$, such that $t_k\geq\tau_k$.
The new number of infected individuals at the end of the day is then obtained as 
\begin{equation}
\label{dinamica}
I(t+\Delta t)=I(t)+n_{inf}(t)-n_{rec}(t).
\end{equation}

\subsubsection{The growth-based basic reproductive ratio \Rc. }
We begin by defining an instantaneous reproductive rate \R(t) as the product of the average number of new infections produced by an infected individual per unit time and the mean duration of infection,
\begin{equation}
\label{Rdet}
\mathfrak{R}(t)= \frac{\langle n_{inf}(t)\rangle}{\langle I(t) \rangle} \frac{\langle \tau \rangle }{\Delta t}.
\end{equation}

Here $\langle n_{inf}(t) \rangle$  and $\langle I(t)\rangle$ denote averages of $ninf(t)$ and $I(t)$ over several simulations, and $\langle \tau \rangle$ is the mean time individuals remain infectious.
Thus, \R$(t)$ represents the number of infections that would be produced per infected individual over the mean infectious period, provided that \R$(t)$ remains constant during this time.
The quantity $\langle \tau \rangle$ equals $\tau_{inf}$ in the case of DR, but for ER it is given by $\langle \tau \rangle=\Delta t / (1-e^{-\Delta t/\tau_{inf}})$, due to the use of a discrete time step in the simulation procedure described in this section (see SM for details). 

With this definition, it follows that, after an initial transient, if the average number of new infections produced daily, $\langle n_{inf}(t) \rangle$, grows exponentially over a given time interval, then \R(t) will be approximately constant over that interval, and we denote this constant by \Rc.
\begin{equation}
\label{R0}
\mathfrak{R}_0 \sim \mathfrak{R}(\epsilon), ~{\rm with~} t_0<\epsilon<t_D.
\end{equation}
Here, $t_0$ denotes the time after which initial transitory effects have subsided, and $t_D$ denotes the time when the depletion of susceptible individuals begins to be noticeable.

Assuming a given type of recovery, an explicit relationship between the exponent $r$ and \Rc\ can be obtained (see SM), which does not depend on the underlying contact structure. This growth-based basic reproductive ratio \Rc\ (which has also been used by other authors \cite{keeling1999,amundsenR0,house_keeling,Kiss2017,fabriciusimmun}) is therefore useful for our study, since if two scenarios with different contact structures have the same \Rc\ (for a given type of recovery), they will exhibit the same growth exponent $r$ at the beginning of the epidemic.
This is not the case for the basic reproductive ratio, $R_0$, defined as the average number of secondary infections that a single infected individual produces in a completely susceptible population \cite{AMlibro, diekmann2000book}. 
Indeed, the relationship between $R_0$ and the growth exponent $r$ requires knowledge of the generation time distribution \cite{wallinga2007gtau}, which depends not only on the intrinsic characteristics of the infectious process within the individual, but also on the local effects of the social structure in which the disease spreads \cite{park2020}. 
Therefore, \Rc\ becomes a convenient quantity to address the main question posed in the Introduction: for an epidemic growing exponentially at a given rate $r$, how does its subsequent evolution depend on the relative weight of frequent versus occasional contacts in the underlying contact structure? To this end, we study scenarios with different contact structures but the same \Rc, which (for a given type of recovery) is equivalent to having the same growth exponent $r$. The advantage of formulating the problem in terms of $r$ lies in the fact that it is a quantity that can be directly estimated from daily reported cases, which are often the only data available in practice.

Finally, it is worth noting that both reproductive ratios, \Rc\ and $R_0$, yield the same epidemic threshold: $R_0$=1 $\iff$ \Rc=1 \cite{Kiss2017}.

\subsection{Deterministic SIR model in discrete time}

In order to evaluate under which conditions the results of our model can be described by the homogeneous mixing approximation, we consider a deterministic SIR model with which to compare our results. 
 Here, we also use 
$S(t)$, $I(t)$, and $R(t)$ to denote the number of susceptible, infected, and recovered individuals in the system at time $t$.
To formulate this model, we start from a stochastic SIR-IBM with a social structure represented by a complete network where all individuals in the system are connected with the same transmission rate $\beta_1=\beta_0/N$. Then, the probability that a susceptible individual will become infected in $\Delta t$ is given by 

\begin{equation}
\label{P1}
P_1=1-e^{-\beta_0 I(t) \Delta t /N}
\end{equation}

\noindent and the number of infections occurring in $\Delta t$  will follow a binomial distribution with mean $P_1 S(t)$. 
Therefore, it is expected that, in the limit of large populations, this IBM will be well represented by a discrete-time deterministic model where the daily infectious rate, $\mathrm{Inc}(t)$, is given by $\mathrm{Inc}(t)=P_1 S(t)$. Concerning recovery from infection, two approximations are considered in our IBM. In the case of ER, the probability that an infected individual will be recovered in $\Delta t$ is given by $1-e^{-\gamma \Delta t}$, with $\gamma=1/\tau_{inf}$, 
so the average number of recoveries during day $t$,  which we take as the recovery rate $\mathrm{Rec}(t)$ in the deterministic model, is given by

\begin{equation}
\label{recE}
{\rm ER:} ~~ \mathrm{Rec}(t)=I(t) (1-e^{-\gamma \Delta t})
\end{equation}

In the case of DR, the number of individuals that recover from infection is the number of individuals that have been previously infected a time $\tau_{inf}$, that is,

\begin{equation}
\label{recD}
{\rm DR:} ~~ \mathrm{Rec}(t)={\rm inf}(t-\tau_{inf}), ~~~ t>\tau_{inf}
\end{equation}

The finite difference equations to be solved are given by:

 \begin{align} 
 \label{deterSIR}
  \begin{aligned}
 S(t+\Delta t)  &= S(t) - \mathrm{Inc}(t) \\
 I(t+\Delta t)  &= I(t) + \mathrm{Inc}(t) - \mathrm{Rec}(t) \\
 R(t+\Delta t)  &= R(t) + \mathrm{Rec}(t) \\
  \end{aligned}
 \end{align}
 
\noindent where  $\mathrm{Inc}(t)=P_1 S(t)$ ($P_1$ from Eq.\ref{P1}), and $\mathrm{Rec}(t)$ is given by Eqs.\ref{recE} and \ref{recD} for ER and DR respectively. For this model, \Rc\ can be obtained from Eq.(\ref{Rdet}) at $t=0$ by taking $n_{inf}(0)=\mathrm{Inc}(0)$.
This leads  to $\RinFla_0=\beta_0 \langle \tau \rangle$  for large $N$, where $\langle \tau \rangle=\tau_{inf}=1/\gamma$ for DR and $\langle \tau \rangle=\Delta t / (1-e^{-\Delta t/\tau_{inf}})$ for ER (see SM). In this last case, if we take the limit $\Delta t \rightarrow 0$, the \Rc\ value $\beta_0/\gamma$ of the classical deterministic SIR model
in continuous time is obtained. 

\vskip 1cm

\section{Results}
\label{results}

In this section, we present the results of simulations for different sets of parameters corresponding to various epidemiological scenarios. 
In all scenarios, we keep the parameters that define the network structure fixed; that is, changes in the contact structure are introduced by varying the transmission rates rather than the connections between individuals themselves.
Regarding recovery from infection, we set $\tau_{inf}=10$ days for both scenarios considered: ER and DR\footnote{
In fact, the value of $\tau_{inf}$ is fairly arbitrary and basically defines the time scale.
If we multiply $\tau_{inf}$ by a given factor, it would suffice to divide the contact rates by the same factor to keep \Rc\ unchanged.
However, it is worth noting that the results would not be strictly identical, since the model dynamics are solved with a finite time step of $\Delta t = 1$ day.
We have verified for the main scenarios that the results do not change qualitatively when setting $\tau_{inf} = 100$ days.
}. 
To assign  values to $\beta_T$ for $T=H, F, W, B,$ and $Q$, we consider several possibilities regarding the proportion of infections acquired through frequent (local) versus occasional (global) contacts. Among these, we selected two paradigmatic cases to serve as baseline scenarios for  our study.
Both scenarios allow the epidemic to spread under certain restrictions, reflecting two distinct epidemic control strategies.
One strategy imposes stricter limitations on going out, such as restrictions on stores, public transportation, and public spaces, leading to more contagion in local settings such as households, families, and workplaces where precautions are relaxed (Scenario L). The other strategy involves lighter restrictions on going out in public spaces, while maintaining stricter precautions in interactions with known individuals, such as stronger enforcement of mask use when meeting coworkers and family members (Scenario G). Accordingly, in Scenario L, disease transmission occurs primarily through local (frequent) contacts, whereas in Scenario G it occurs mainly through global (occasional) contacts.
The parameters for both scenarios have been determined considering that, in both cases,  the same value of $\RinFla_0 \simeq 1.5$ is obtained. This is a plausible value, for example, for a COVID-19 epidemic spreading under restrictions \cite{fabriciusimmun}.
For the computation of \Rc\ through Eqs. \ref{R0} and \ref{Rdet}, we performed simulations using $I_{ini}=0.02\%$ and averaged the results over 50 simulations for each scenario.
 We choose this relatively high value of $I_{ini}$ to reduce the shift observed between curves corresponding to different stochastic realizations (see Ref.~\cite{Kiss2017}, p.~387). Otherwise, the plateau of \R$(t)$ is not well defined, making it difficult to determine \Rc.

The values of $\beta_T$ defining scenarios L and G, and the corresponding $\beta_l$ values derived from them,  for different locations, are provided in Table \ref{ratesLandG2}. The daily contact probability between individuals, $p_l^C$, is also presented.
The $p_l^C$-values help us to quantify our classification of contacts 
as frequent or occasional. To estimate the probability of having
a contact at location $l$ over a time period $\Delta t = 1 $ day, we assumed $\beta_l = c_t \beta^C_l$, where $\beta^C_l$ is the contact rate and $c_t$ is the probability of transmission given contact\footnote{Decomposing the transmission rate into a contact rate and a per-contact transmission probability is a standard modeling simplification \cite{wallinga2006}. 
However, we formulate the model directly in terms of an effective transmission rate $\beta_T$ for each location type $T$, since the per-contact transmission probability, $c_t$, depends strongly on the specific disease, and our study does not focus on any particular one.}

\begin{equation} \label{pC}
p^C_l(\Delta t) = 1 - e^{-\beta^C_l \Delta t}
\end{equation}

To make this estimation, we used $c_t=0.03$, based on estimates for COVID-19 in different countries \cite{pnaswilder}. It is worth noting that this value of $c_t$ does not affect our calculations since the parameters that enter the simulation are the transmission rates $\beta_l$ defined from the $\beta_T$. However, including an approximate estimate of $c_t$ allows us to make a rough estimation of $p^C_l$. As can be seen in the table, for the contacts identified as frequent, the daily probability of encountering is on the order of 1, while for occasional contacts, $p^C_l \ll 1$. 
For all the simulations carried out in this section, we set $I_{ini}=36$.

Table \ref{infections} presents the total number of infections occurring in each type of location for L and G scenarios and the two types of recovery. As can be observed, in both scenarios, the majority of infections occur through frequent contacts. However, while in G scenarios, 40\% of infections occur through occasional contacts, in L scenarios, this figure does not exceed 10\%.
Note that for DR, we slightly modify the $\beta_T$ values of Table \ref{ratesLandG2}  to obtain the same \Rc\ (within our numerical precision) and similar proportions of infections that occur through frequent versus occasional contacts as for ER. The values of the parameters for DR are shown in Table S1 (see SM). 
 
Fig.\ref{Figure1} shows the evolution of the epidemic in both scenarios for the two types of recovery from infection considered. As can be seen, the epidemic develops faster for scenario L, the prevalence reaches a higher value at the peak, and the fraction of the population finally infected is also higher.
However, it is important to note that these differences are much more pronounced in the case of DR. 

Next, we considered implementing a relaxation measure allowing moderate school attendance. By moderate, we mean that only within-class and small-group interactions ($Sch_1$ and $Sch_2$) are included in the contact structure, while interactions between individuals from different classes are excluded. To compare the impact of the effect in both scenarios, we considered the same transmission rates for the interactions introduced in each scenario ($\beta_{Sch_1}=0.062$ 1/day and $\beta_{Sch_2}=0.02$ 1/day) for both recovery types. Fig.\ref{Figure2} shows that this relaxation measure has a greater impact on scenario L than on scenario G. In the case of ER, the peak height of $i(t)$ in scenario L increases by 90\% 
when class attendance is included, while in scenario G, the increase is 63\% 
(left top panel). For DR, the increases in peak height are 64\% 
and 57\%
for scenarios L and G, respectively (right top panel). The values of \Rc\ also show a greater increase for scenario L compared to scenario G (bottom panels). 
However, the relative increase in the number of individuals affected by the epidemic is lower in both the L and G scenarios, and it always remains below 17\%.

On the other hand, if the 
worsening of the epidemiological situation involves an equal increase in the transmission rate  for all individuals—potentially caused by a generalized removal of face masks, for instance, or the circulation of a pathogen with higher infectivity—a greater impact is observed for G scenarios compared to L scenarios. To simulate this situation, we multiplied all the transmission rates, $\beta_l$, by the same factor. First, we considered an increase in the rates to achieve an effect of similar magnitude to that obtained in the previous case (inclusion of moderate school attendance). To do this, we multiplied the rates by a factor  $f= (\RinFla_0^{(L+Sch)} + \RinFla_0^{(G+Sch)}) / (\RinFla_0^{(L)} + \RinFla_0^{(G)})$ that gives 1.24 
for ER and 1.20 
for DR (the \Rc-values are given in the captions of Figs. \ref{Figure1} and \ref{Figure2}). Fig.\ref{Figure3} shows the curves obtained for $i(t)$ and \R$(t)$.
For both types of recovery, the change in transmission rates leads to a greater increase in the peak height of $i(t)$ for the G scenario (97\% for ER and 76\% for DR) than for the L scenario (89\% for ER and 56\% for DR), while nearly the same value of  \Rc\ is obtained for both scenarios.
If the rates are multiplied by a factor of 2, the trend observed in the previous case becomes more pronounced, and the $i(t)$ curves obtained for the L and G scenarios become very similar for both types of recovery (Fig.~\ref{Figuredoubling}). 
On the other hand, this is the first time that higher values of \Rc\ are observed for the G scenarios compared to the L ones.

\section{Discussion}
\label{discuss}
\begin{figure}[H]
\centering
\includegraphics[width=0.475\columnwidth]{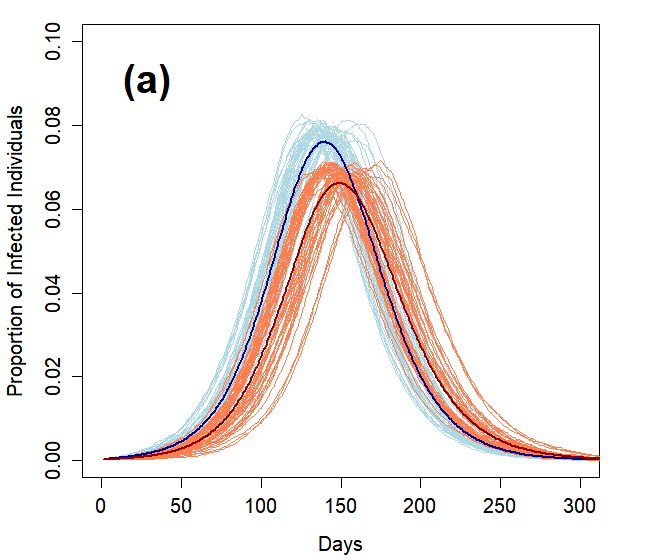}  
\includegraphics[width=0.475\columnwidth]{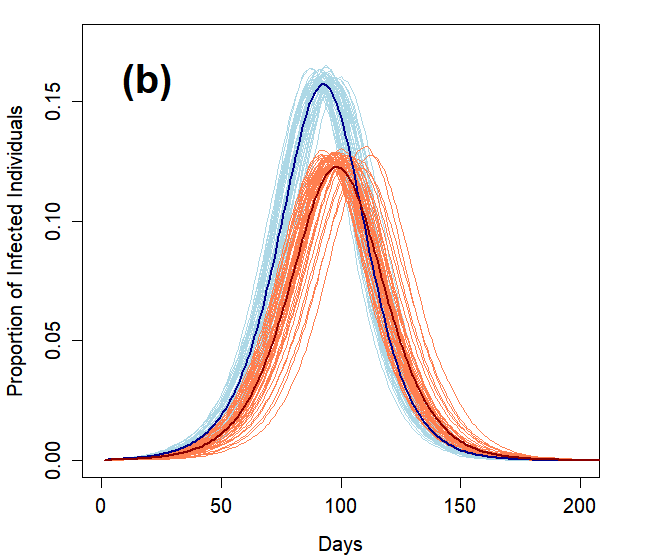}   
\caption{\textbf{Effect of contact structure on epidemic spread.}
Fraction of infected individuals in the system as a function of time for scenarios L (blue and light blue) and G (dark red and orange). 
The lighter colored curves represent 50 simulations, while the darker curve its the average over these results. 
(a) ER. ($ \RinFla_0^{(G)} = 1.51 \pm 0.04, 
\RinFla_0^{(L)} = 1.51 \pm 0.03$).
Scenario L had 9\%
more infected individuals by the end of the epidemic, and 14\%
more cases at the peak.
(b) DR. ($ \RinFla_0^{(G)} = 1.51 \pm 0.04,
\RinFla_0^{(L)} = 1.51 \pm 0.04$)
Scenario L had 20\% more infected individuals by the end of the epidemic, and 27\% more cases at the peak.
}
\label{Figure1}
\end{figure}

\begin{figure}[h!]
\centering
\includegraphics[width=0.475\columnwidth, width=0.425\linewidth]{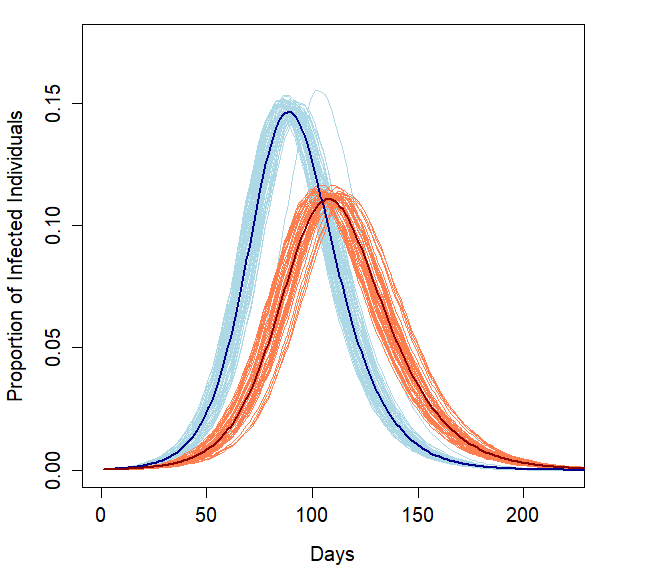}  
\includegraphics[width=0.475\columnwidth, width=0.425\linewidth]{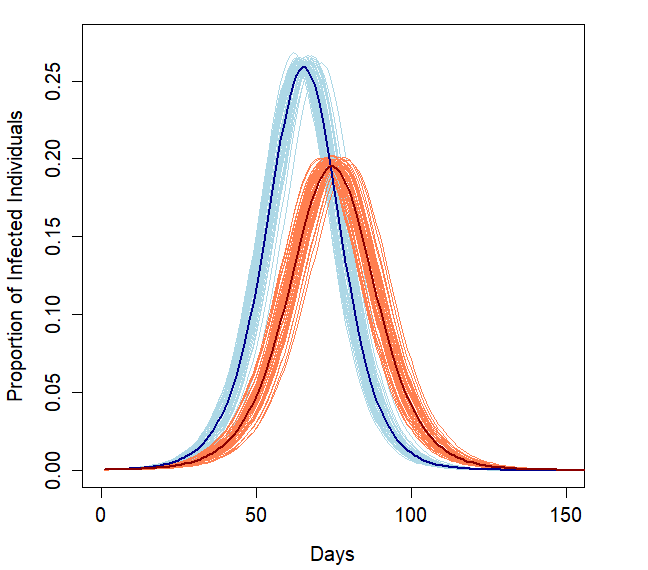} 
\includegraphics[width=0.475\columnwidth, width=0.425\linewidth]{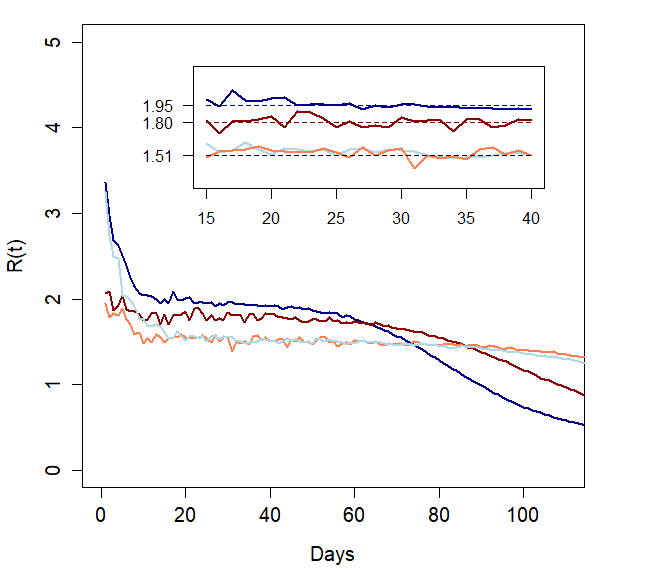} 
\includegraphics[width=0.475\columnwidth, width=0.425\linewidth]{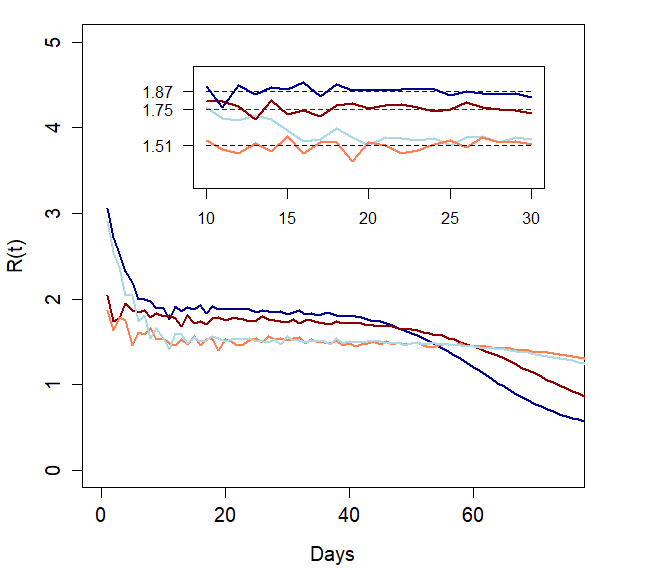}
\caption{\textbf{Effect of including moderate school attendance on epidemic spread}.
Top panels: Same as Fig.\ref{Figure1} when within-class and small-group interactions ($Sch_1$ and $Sch_2$) are included in the contact structure. Bottom panels: Reproductive ratio, \R(t), for L and G scenarios before and after incorporating school interactions; all curves represent averages over 50 samples; light blue and coral correspond to L and G scenarios, respectively, before adding school interactions, while dark blue and dark red represent the scenarios after including school interactions. Insets: detail of \R(t) in the time interval used to compute \Rc\ in the corresponding cases. Left panels: ER. After including school interactions: $ \RinFla_0^{(G+Sch)} = 1.80 \pm 0.06, \RinFla_0^{(L+Sch)} = 1.95 \pm 0.05$. Right panels: DR. After including school interactions: $ \RinFla_0^{(G+Sch)} = 1.75 \pm 0.06, \RinFla_0^{(L+Sch)} = 1.87 \pm 0.05$). In both cases \Rc\ values before including school interactions are given in Fig.\ref{Figure1}.
}

\label{Figure2}
\end{figure}

\begin{figure}[h!]
\centering
\includegraphics[width=0.475\columnwidth, width=0.425\linewidth]{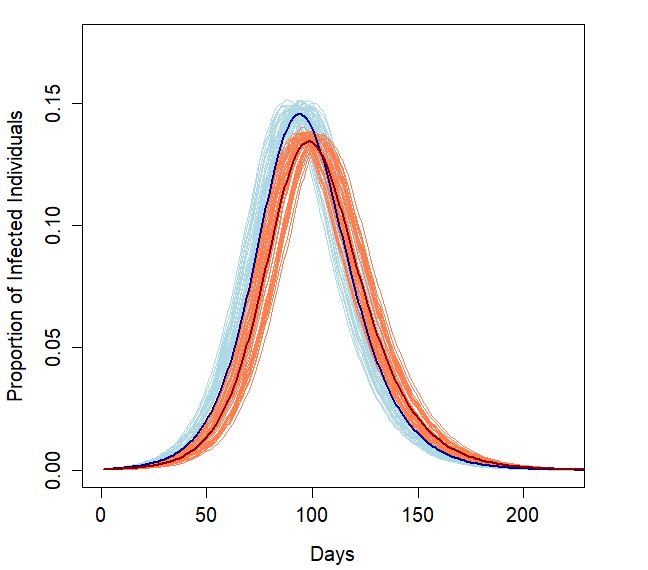}  
\includegraphics[width=0.475\columnwidth, width=0.425\linewidth]{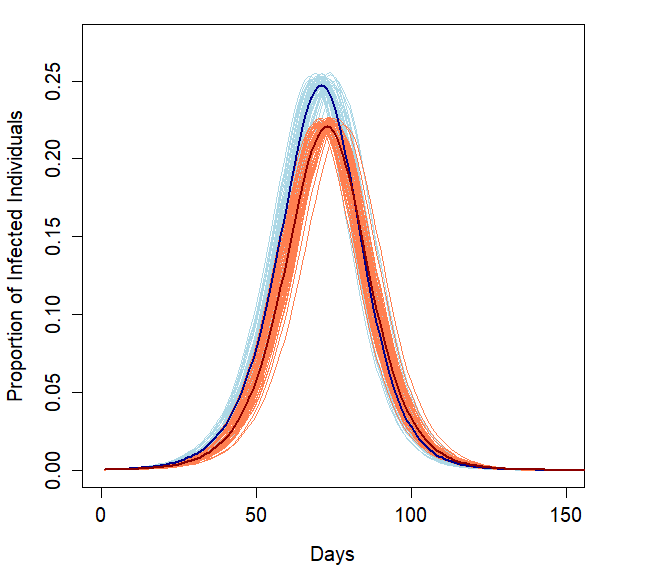} 
\includegraphics[width=0.475\columnwidth, width=0.425\linewidth]{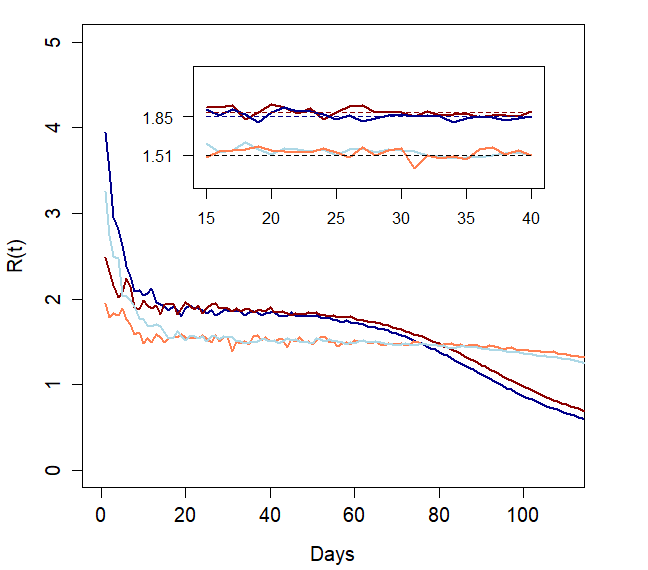}  
\includegraphics[width=0.475\columnwidth, width=0.425\linewidth]{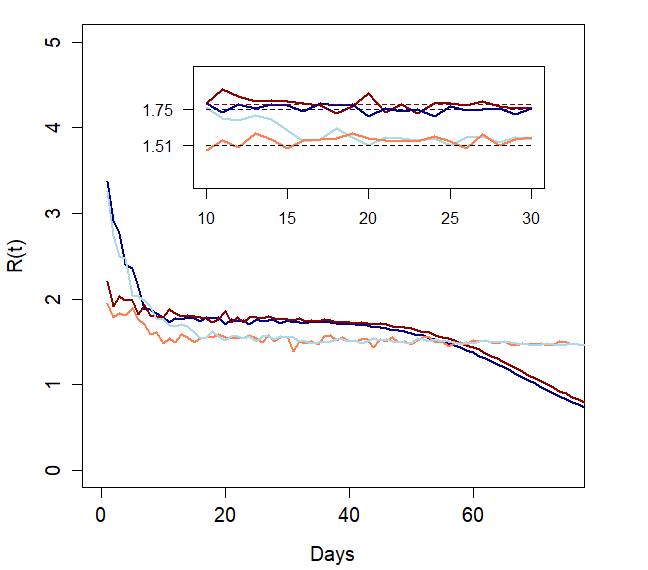}
\caption{\textbf{Effect of uniformly increasing all transmission rates between individuals on epidemic spread.} Top: Infected curves over time for the scenarios L (blue and light blue) and G (dark red and orange) when all transmission rates are increased by a factor $f$ ($f=1.24$ for ER, and $f=1.20$ for DR).
The lighter colored curves represent 50 simulations of the epidemic curves, while the darker curve is the average over these results. Bottom: Reproductive Ratio, \R(t), as a function of time for scenarios L and G before (light blue and coral respectively) and after (dark blue/dark red) applying the multiplicative factor, $f$; this relaxation measure resulted in $ \RinFla_0^{(G)} = 1.89 \pm 0.05,
\RinFla_0^{(L)} = 1.85 \pm 0.05$ for ER (left panel) and
$ \RinFla_0^{(G)} = 1.78 \pm 0.06, \RinFla_0^{(L)} = 1.75 \pm 0.06$ for DR (right panel). }

\label{Figure3}
\end{figure}

\begin{figure}[h!]
\centering
\includegraphics[width=0.475\columnwidth, width=0.425\linewidth]{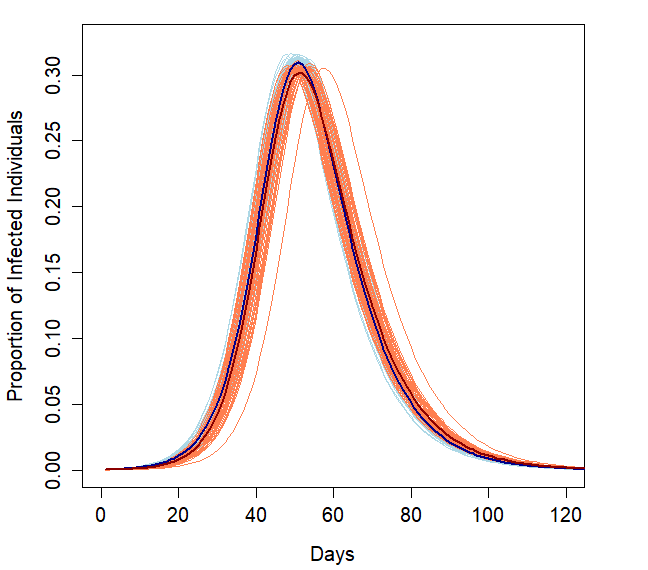}  
\includegraphics[width=0.475\columnwidth, width=0.425\linewidth]{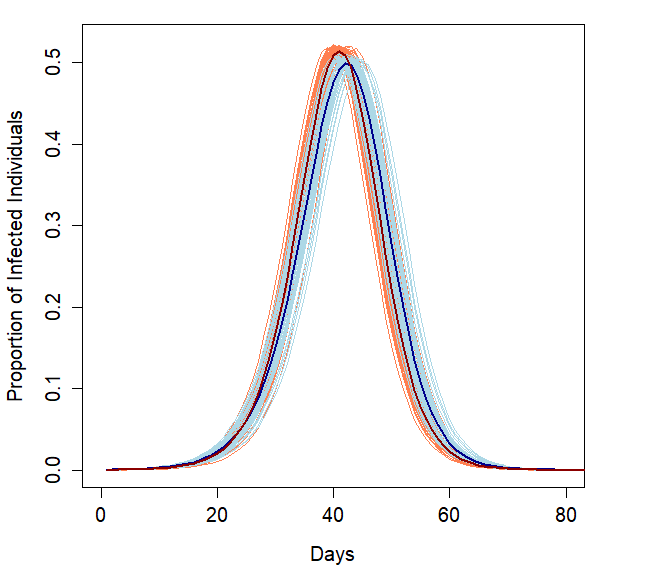} 
\includegraphics[width=0.475\columnwidth, width=0.425\linewidth]{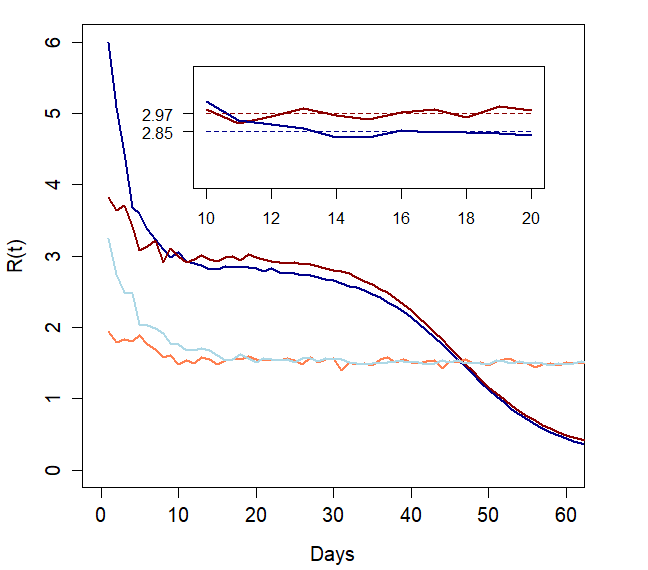}  
\includegraphics[width=0.475\columnwidth, width=0.425\linewidth]{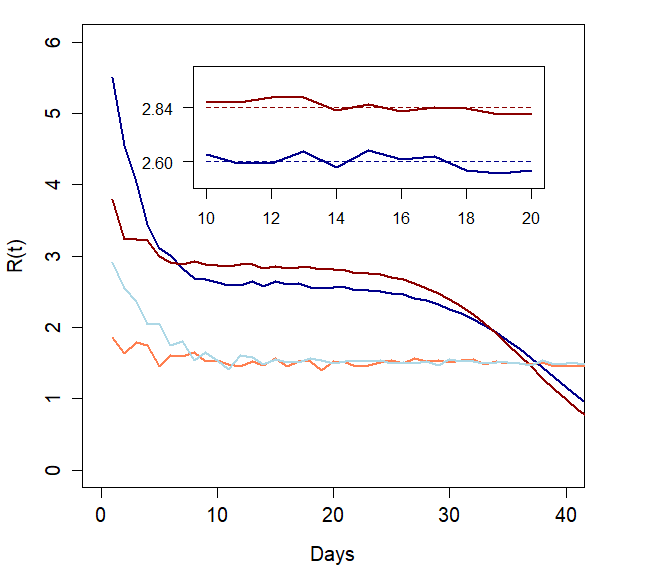}
\caption{\textbf{Effect of doubling transmission rates on epidemic spread.} Top: Infected curves over time for scenarios L (blue and light blue) and G (dark red and orange) 
when all transmission rates are increased by a factor $f=2$. The lighter colored curves represent 50 simulations of the epidemic curves, while the darker curve is the average over these results. Bottom: Reproductive ratio, \R(t), as a function of time for scenarios L and G before (light blue and coral respectively) and after (dark blue/dark red) applying the multiplicative factor; this relaxation measure resulted in $ \RinFla_0^{(G)} = 2.97 \pm 0.10, \RinFla_0^{(L)} = 2.85 \pm 0.09$ for ER (left panel) and $ \RinFla_0^{(G)} = 2.84 \pm 0.09, \RinFla_0^{(L)} = 2.60 \pm 0.08$ for DR (right panel). }

\label{Figuredoubling}
\end{figure}

 In this section, we analyze our results and present additional findings that provide further insight into specific aspects.
In this work, we conducted our study considering two different types of recovery from infection: exponentially distributed recovery times (ER) and fixed recovery times (deterministic recovery: DR). It is important to note that epidemic dynamics can differ significantly even when sharing the same \Rc\ under these two recovery types. 
As is well known, an epidemic with deterministic recovery (DR) spreads with a higher exponent, reaching its peak sooner and with greater height \cite{KRlibro}. These characteristics become evident when comparing the axis scales in Figs.\ref{Figure1}a and \ref{Figure1}b. However, many differences in the spreading dynamics between scenarios L and G remain qualitatively similar for both types of recovery (Figs.\ref{Figure1}-\ref{Figuredoubling}).

\begin{figure}[h!]
\centering
\includegraphics[width=0.475\columnwidth, width=0.425\linewidth]{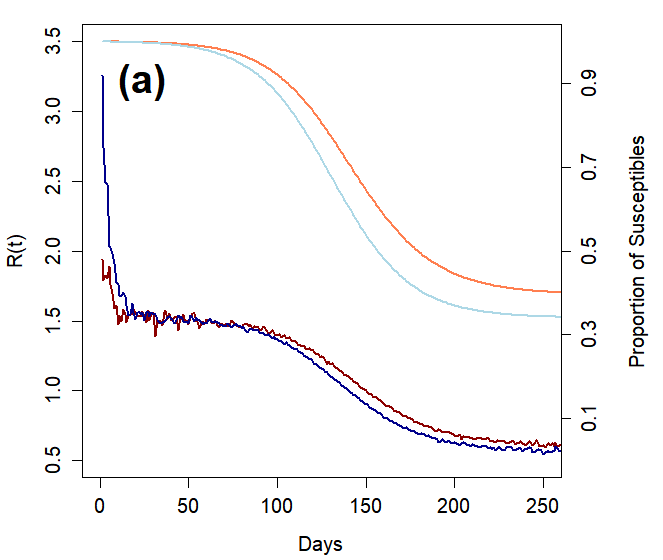}  
\includegraphics[width=0.475\columnwidth, width=0.425\linewidth]{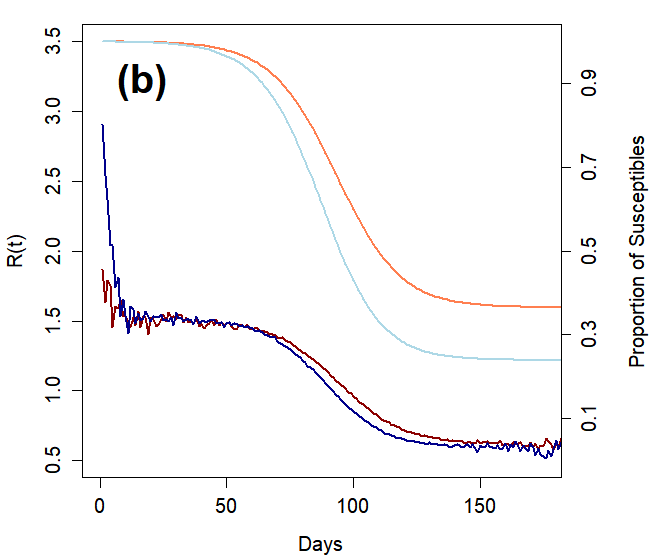} 
\includegraphics[width=0.475\columnwidth, width=0.425\linewidth]{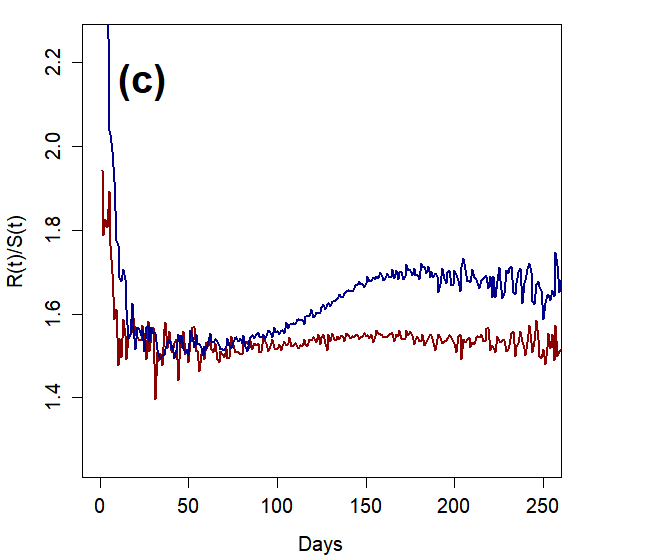}  
\includegraphics[width=0.475\columnwidth, width=0.425\linewidth]{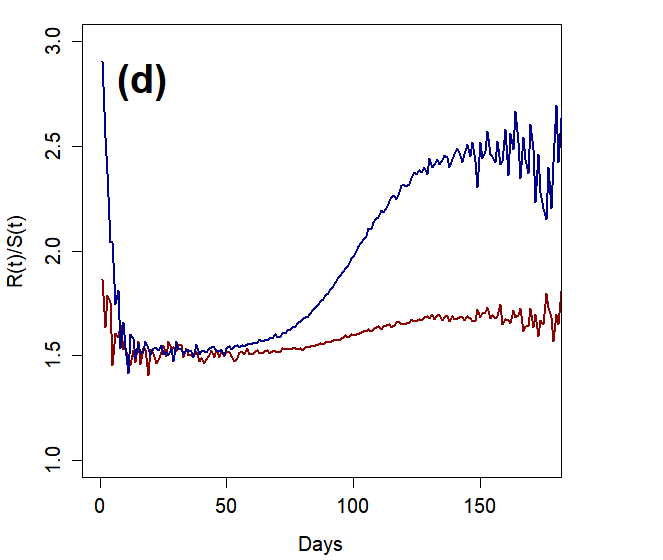}

\caption{\textbf{Relationship between the reproductive ratio and the availability of susceptible individuals.}
(a) Reproductive ratio,\R(t) , for scenarios L and G (dark blue and dark red, respectively; left y-axis), and fraction of susceptible individuals, $s(t)$, for scenarios L and G (light blue and coral, respectively; right y-axis), for the case of ER.
(b) Same as (a), but for DR.
(c) Ratio \R$(t) /s(t)$ for scenarios L and G (dark blue and dark red, respectively) in the case of ER.
(d) Same as (c), but for DR.
}
\label{efecto}
\end{figure}

\begin{figure}[h!]
\centering
\includegraphics[width=0.475\columnwidth, width=0.425\linewidth]{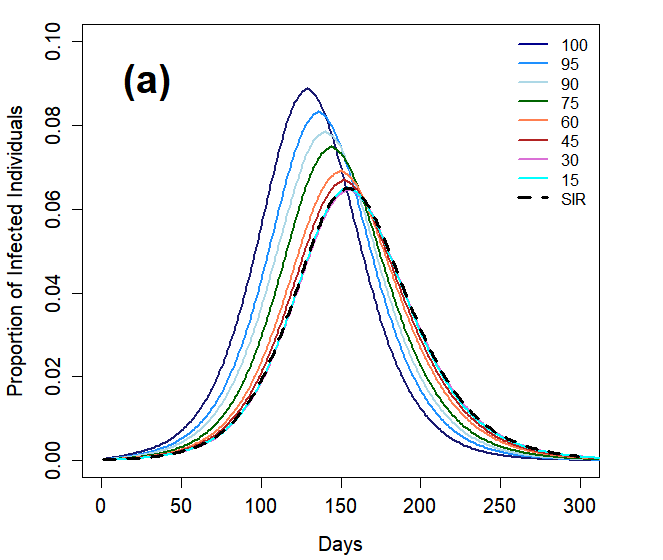}  
\includegraphics[width=0.475\columnwidth, width=0.425\linewidth]{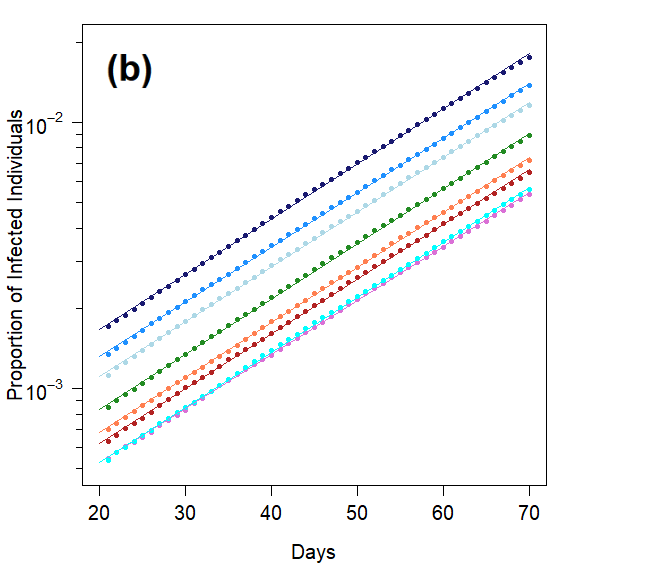}  
\includegraphics[width=0.475\columnwidth, width=0.425\linewidth]{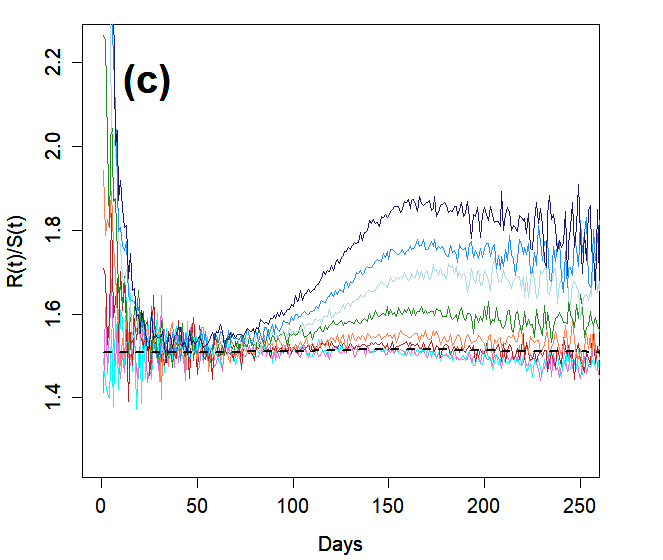}
\includegraphics[width=0.475\columnwidth, width=0.425\linewidth]{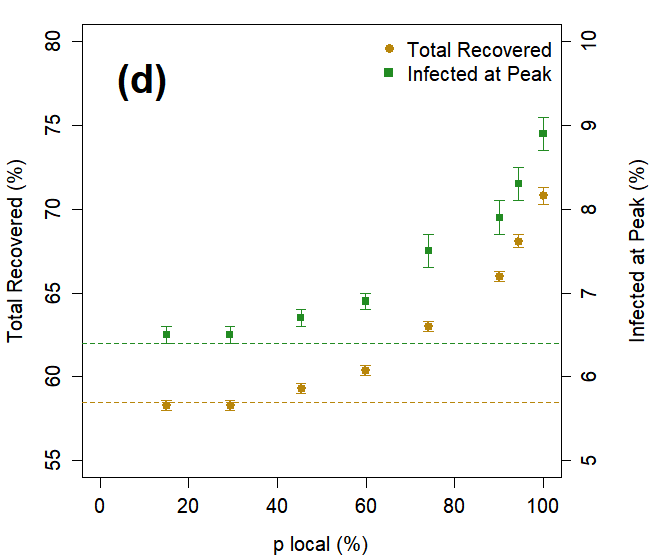}
  
\caption{\textbf{
Epidemic spread in scenarios with different proportions of \textit{local} contacts.}
 System dynamics for scenarios where 15\% to 100\% of infections
 are produced through frequent contacts (at locations $H, F$ and $W$) for ER. The colored curves represent the average over 50 simulations. 
 Deterministic SIR model in discrete time (black dashed line). 
 (a) Fraction of infected individuals as a function of time. In all cases $\RinFla_0 \sim 1.51$, so we take $\RinFla_0^{SIR} = 1.51$. 
 (b) Same as in Figure (a) for the start of epidemic spread. The results of the simulations (points) and an exponential fit (lines) are shown in logarithmic scale.
 (c) Ratio between reproductive ratio and fraction of susceptible individuals, \R$(t)/s(t)$.
 (d) Fraction of the population infected at the peak (yellow points) and total fraction of the population infected (green points) 
 as a function of the percentage $p$ of infections that
 are produced through frequent contacts. Dotted lines are HM values for \Rc=1.51.
 }
\label{porcilocal}
\end{figure}

Undoubtedly, one of the most interesting findings of our study is that, in the L scenario, the epidemic has a greater impact than in the G scenario—this feature being more pronounced in the case of deterministic recovery (Fig.\ref{Figure1}).
Let us now examine how this effect can be explained.
It is a well-established fact that, when local interactions occur, there is a clustering and screening of infected individuals, which reduces transmission \cite{KRlibro, keeling1999, eames2008, simoes, sirsw}. Frequent contacts with the same individuals cease to be effective in spreading the infection once that close environment has already been infected. Therefore, for a given average number of contacts per individual, the epidemic grows faster if the contacts occur with different individuals \cite{keeling1999, eames2008}. What is particular about our approach is that we compare scenarios with greater and lesser weight of local interactions, while ensuring that they have the same reproduction rate, \R(t), once the initial transient phase has passed. This value of  \R(t), which remains stable until the depletion of susceptibles begins, is what we have defined as  \Rc.
 To achieve the same \Rc\ value in both the L and G scenarios, individuals in the L case must have a higher average number of contacts per individual, which initially (when the entire population is susceptible) leads to a greater number of infections. However, the aforementioned screening effect causes a sharp drop in \R(t) during the early days of the epidemic (Fig.\ref{efecto}a, \ref{efecto}c). 
On day 15, once the transient phase has passed and the epidemic begins to spread with the same reproduction rate in both scenarios ($\RinFla_0 \sim 1.5$), the L scenario shows only a slightly higher number of infected individuals than the G scenario—around 60 more in the case of ER and 80 more in the case of DR. This difference, inherited from the initial transient  phase, is enough for the epidemic to start earlier in the L scenario compared to the G scenario (Fig.~\ref{Figure1}). However, it does not account for the higher epidemic peak observed in the L scenario, nor for the significantly larger final number of infections—approximately 10,000 more in the case of ER and 22,700 more in the case of DR.
In Figures \ref{efecto}a and \ref{efecto}c it can be observed that  \R(t) begins to decline when the availability of susceptibles decreases. However, while in G scenario  \R(t) decreases almost proportionally to the fraction of susceptibles in the system,  $s(t)$, in L scenario the decline in \R(t) is smaller (Figs.\ref{efecto}b, \ref{efecto}d) which allows the epidemic in L to have a greater reach. This effect could be explained by the same clustering phenomenon of infected individuals that initially slows the spread of the epidemic. Once these infected individuals recover, they remain grouped in locations where the epidemic has already ``passed'', so the infected individuals colonizing new locations encounter fewer recovered individuals  (on average) than they would if the recovered individuals were distributed randomly.
This correlation effect becomes more pronounced when scenarios with a higher proportion of local interactions are considered, as shown in Figures~\ref{porcilocal}a and~\ref{porcilocal}c for the case of ER. On the other hand, when the weight of local interactions falls below 30\%, this effect disappears entirely, and the system’s dynamics can be accurately described by the homogeneous mixing (HM) approximation (dashed black curve).
To obtain the results presented in Figure 6, we defined a set of scenarios in which the percentage, $p$, of infections occurring through frequent contacts varies between 15\% and 100\%. The transmission rates, $\beta_T$, for these scenarios were chosen so that the fraction of infections occurring in each location type $T$ varies approximately linearly with $p$. The values of $\beta_T$ and the corresponding percentages of infections occurring in each type of location are shown in Tables S1 and S2, respectively (see SM).
 In Fig.\ref{porcilocal}b, we verify that all the curves for the fraction of infected individuals grow exponentially with 
 a similar exponent $r$, which is approximated by Eq. (S11) 
 with a precision better than 1\% (see SM)
 \footnote{We have verified that \Rc\ and $r$ are related by Eq. (S11) with similar precision for all the systems studied in this work.}.
 The differences in the epidemic developments are due to the unequal impact of the depletion of susceptibles as a function of the degree of locality of the interactions (Fig. \ref{porcilocal}c). In the limiting case of 100\% local interactions, the total number of individuals affected by the epidemic and the maximum value of the fraction of infected individuals are approximately  22\% and 40\% higher, respectively, than predicted by the HM approximation (Fig.\ref{porcilocal}d).
As can be seen by comparing Figures \ref{Figure1}, \ref{efecto}, \ref{porcilocal}, and S2, this correlation effect is much more pronounced in the case of deterministic recovery. This is due to the lower probability that an infected individual will recover before transmitting the infection to someone in a location where all others are still susceptible (see SM, Section S6). 
As a consequence, the transmission network established among locations H, W, and F-and therefore the resulting correlations-becomes stronger under deterministic recovery (DR).

\begin{figure}[h!]
\centering
\includegraphics[width=0.475\columnwidth]{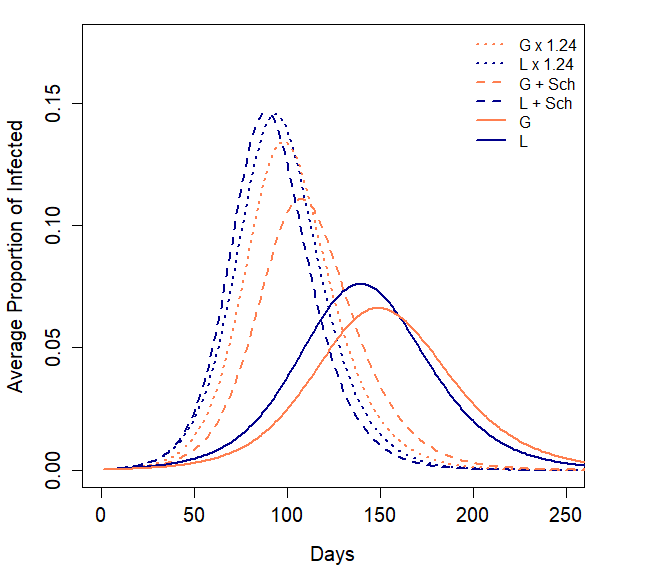}
\includegraphics[width=0.475\columnwidth]{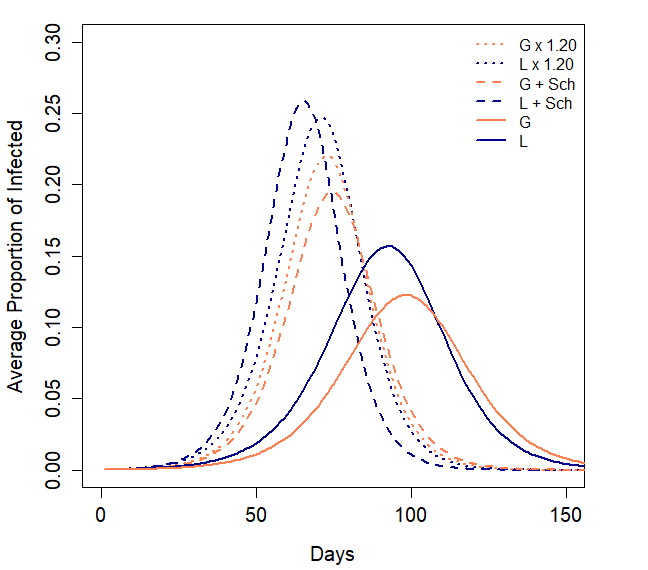}
\caption{\textbf{Comparison of the effects
of including moderate school attendance
and
uniformly increasing all transmission rates for L and G scenarios.}
Continuous, dashed, and dotted curves correspond to the averages over the samples from Figures 1, 2, and 3, respectively, which are plotted together here for better comparison. Left panel: ER; right panel: DR.
}
\label{FigureSyf}
\end{figure}

To facilitate the comparison of the impact of the relaxation effects on scenarios L and G (Figs.\ref{Figure2} and \ref{Figure3}), Fig.\ref{FigureSyf} presents the $i(t)$ curves for cases L and G before and after introducing changes in the social structure. It can be observed that while the introduction of moderate school attendance increases the differences in the propagation dynamics between L and G, multiplying all the rates by the same factor reduces them. In particular, if the factor continues to increase, the $i(t)$ curves of both scenarios become very similar (Fig.\ref{Figuredoubling}). 
The greater relative increase in the $i(t)$ curve for the G scenario compared to the L scenario is due to the fact that, when transmission rates are already high, further increasing them does not significantly raise transmission in local settings—as it does in global ones—because of the saturation of infections that occurs in the former, particularly within households.

In Figures 6 and S2, we saw that when the weight of local interactions is 30\%, the homogeneous mixing (HM) approximation already provides a good description of our epidemic model. However, for both the L and G scenarios, HM still underestimates the values of $i(t)$. To extend this comparison to the other cases studied, in Figure S3 (see SM) we compare the curves from the 50 simulations performed for each case (previously presented in Figures 1, 2, 3, and 4) with the corresponding HM prediction. It can be observed that the agreement between the HM prediction and the agent-based model curves for $i(t)$ improves as the value of $R_0$ increases.
However, it should be noted that in the case where moderate school attendance is included, although the SIR model provides a reasonable approximation of the 
$i(t)$ curves for each scenario when the appropriate 
\Rc\ value is supplied, the homogeneous mixing (HM) approximation cannot provide a prediction for \Rc\ in this case, which, in fact, differs between the L and G scenarios.

Although the parameters chosen to define scenarios L and G do not represent the social structure of any particular region, their values were chosen to approach a plausible daily number of contacts per individual at the most common locations (home, work, etc.), in agreement with reports from the literature. For example, in the POLYMOD study \cite{mossong2008}, where extensive surveys were conducted in eight European countries, the distribution of all pooled reported contacts was as follows: 23\% at home, 21\% at work, 14\% at school, 3\% while traveling, and 16\% during leisure activities. Of the remaining 23\%, 15\% corresponded to various other locations, and 8\% reported contacts with the same person in multiple locations. 
To estimate the percentage of contacts that occur in each location in our model, we assume—as in Section \ref{results}—that there is a probability of transmission given contact ($c_t$), regardless of where the contact takes place. However, to estimate these percentage values, it is not necessary to assume a specific value for $c_t$, unlike in Section \ref{results}.
For G scenario, we found that 40\%, 6\%, 16\%, 8\%, 7\%, and 23\% of contacts occur at $H, F, W, S, B$, and $Q$ locations, respectively (see SM, Section S4). When comparing our results with those of POLYMOD, the following considerations should be taken into account:
\textit{(i)} contacts occurring at our F location are likely to take place at home, which would result in 46\% of contacts happening at home; \textit{(ii)} in POLYMOD, contacts categorized as occurring while traveling, during leisure activities, and at other places sum to 34\%, which corresponds to our $B$ and $Q$ locations; \textit{(iii)} our study represents a social dynamic under restrictions, whereas POLYMOD does not. Regarding this last point, it is not obvious how the POLYMOD results would be modified when considering restrictions. For example, if the number of contacts at work and school decreases, this would likely lead to some increase in home contacts. For the purpose of comparing with scenario G, we assume that the restrictions leading to this scenario would reduce contacts by half in all POLYMOD settings except at home, and we further assume that half of the contacts reduced from the workplace occur at home (due to remote work). A renormalization would yield 45\%, 17\%, 11\%, and 27\% of contacts at H, W, S, and other locations (those mentioned in \textit{(ii)}) that compare well with the 46\%, 16\%, 8\%, and 30\% observed in our study for the G scenario.

Finally, we have verified that the central result of this work—namely, that for \Rc=1.5 the overall epidemic size increases with the relative weight of local contacts in transmission—does not depend on the specific features of the network structure employed.
 To this end, we analyzed an extremely simple contact structure consisting only of households and workplaces, with two population groups, only one of which participates in the workforce. 
In fact, one could argue that this structure cannot be further simplified, since a system composed solely of isolated households would not allow any transmission.
For this minimal structure, where transmission occurs exclusively through local contacts, we find that the total number of infected individuals is 37\% and 63\% higher than in the homogeneous mixing (HM) case, for ER and DR, respectively (in all cases for the same \Rc=1.5; see Section~S8 of the SM for details).

\section{Conclusions}
\label{conc}
In this study, we used a stochastic individual-based model to analyze the spread of an infectious disease within a social structure representing a context in which restrictions and precautions shape social behavior.
The key finding of our work is that, for epidemics exhibiting the same initial growth rate, the eventual impact is larger the greater the contribution of frequent contacts (i.e., local settings) to transmission.
In models of this kind, where there is a certain degree of arbitrariness in both the structural details and parameter choices, conclusions formally apply to the specific system under study. However, it is worth noting that the baseline parameter values considered here are epidemiologically plausible, and the main features of the proposed contact structure reflect fairly general patterns. Moreover, we have shown that the key finding persists even in a minimal contact structure including only households and workplaces, suggesting that this effect is general and not a consequence of the specific features introduced by increasing model complexity.
To the best of our knowledge, this result has not been previously reported and carries relevant epidemiological implications. In particular, if interventions are implemented to reduce occasional contacts, care should be taken to avoid a concurrent relaxation of precautions in frequent-contact settings, as this could lead to outcomes contrary to those intended.
Our approach consisted of analyzing different scenarios in which parameters were chosen so that the epidemic initially spreads at the same rate in all cases, thus enabling a direct assessment of how the relative contribution of frequent and occasional contacts shapes the subsequent dynamics.
Interestingly, although the extreme case of a social structure without occasional contacts may be unrealistic, our results show that even under such conditions—where contacts occur only among acquaintances—a widespread epidemic can still emerge (see Fig.\ref{porcilocal}).

In this study, we consider two extreme cases for the distribution of recovery times: exponential and deterministic recovery. 
The fact that the aforementioned conclusions hold for both types of recovery suggests that they are independent of the recovery time distribution. However, it should be emphasized that the effect whereby the epidemic has a greater impact in a scenario dominated by local interactions (for the same initial transmission rate, \Rc) is much more pronounced in the case of deterministic recovery.

Another important conclusion of our study is that the implementation of relaxation measures on a socially restricted setting can have markedly different impacts depending on the characteristics of the transmission scenario in which they are applied. For instance, we found that allowing moderate school attendance has a much greater impact in scenarios where local interactions dominate, whereas a relaxation measure consisting of a uniform increase in the transmission rate for all individuals has a greater impact in scenarios where global settings predominate.

In this study, we also set out a methodological goal: to explore under which conditions the homogeneous mixing (HM) approximation provides a reliable description of our epidemic model. For the baseline case of \Rc\ $\simeq 1.5$ considered here, and for the type of occasional interactions modeled (contacts in gyms, small shops, large stores, public transport, etc., without accounting for superspreading events or
pronounced heterogeneity in contact intensity across individuals), we found that the HM approximation yields a satisfactory description when the weight of such interactions approaches 70\%. 
However, when relaxation measures are implemented and \Rc\ increases to around 2.0, we found that the HM approximation remains valid even when the weight of global interactions drops to 40\%.
This indicates that, for the type of global interactions considered here (i.e., in the absence of strong heterogeneities across individuals), increasing the overall intensity of interactions makes the homogeneous mixing approximation more accurate.

\section*{Acknowledgements}

We acknowledge financial support from the Consejo Nacional de Investigaciones Científicas y Técnicas (CONICET) through PIP 0266 (2022-2024).  G.F. is member of the Scientific Career of Consejo Nacional de Investigaciones Científicas y Tecnológicas-CONICET (Argentina). 

\section*{Declarations}
\subsection*{Conflict of interest}

The authors declare that they have no conflict of interest.

\subsection*{Data availability}
The computational codes in R 
are provided by the authors upon request.

\bibliography{bibs}

\newpage
\begin{table}[H]
\centering
\caption{\label{enes}
\textbf{Type and size of the different locations considered in the model.}
The number of individuals that can be contacted by a specific individual, $N_l$ is equal to  $size$-1 in the case of \textit{places} since the contacting individual also belongs to the location. On the other hand, the ``Extended Family'' does not include the contactor's household, so in this case, $N_l=size$. For details see the Supplementary Material.
}
\begin{tabular}{llc}
\hline
 \multicolumn{2}{c}{Location-type ($T$)}  & $Size$   \\      
\hline
\multirow{6}{2.8em}{Places} &Household ($H$)   & 1, 2 and 4 \\
&Work    ($W$)   &  1 to 17 \\
&School-Classroom  ($Sch_1$)   &  25 \\
&School-Group  ($Sch_2)$   & 8 and 9  \\
&Retail Stores   ($B$)   &  300  \\
&Quarter  ($Q$)  &  $\sim$ 45,000  \\
 \multicolumn{2}{l}{Extended Family ($F$)} & 2, 3, 4, and 8 \\
\hline
\end{tabular}
\end{table}

\begin{table}[H]
\centering
\caption{\label{ratesLandG2} 
  \textbf{Transmission rates for frequent and occasional contacts.}
  Parameters that characterize the contacts among individuals at the different locations for L and G scenarios with ER.
  Locations $l$ can be of different types $T$ = $H, F, W, B$ and $Q$,
  $\beta_l$ is the
     transmission rate per link at location $l$, given by Eq.\ref{betal} in terms of $\beta_T$ and $N_l$, where
      $N_l = \text{Size} - 1 $ is  the number of different individuals that could be contacted at a given location (the location sizes are given in Table \ref{enes}). 
    Rates $\beta_T$ and $\beta_l$ are in 1/day. 
$p^C_l \equiv p^C_l(\Delta t =1)$ is the probability of having a contact with one specific individual of location $l$  during a day, assuming a probability of infection given contact of 0.03. The fact that $p^C_l$ is on the order of unity for locations of type H, F, and W, and much smaller for $B$ and $Q$, allows us to identify contacts that occur in $H, F,$ and $W$ as frequent, and the others as occasional. As for locations $F$ and $W$ different $N_l$ values are possible, minimum and maximum values are indicated in the table for $\beta_l$ and  $p^C_l$. For the case of stores (B), the reported $\beta_l$ value corresponds to the case in which individuals frequent only one store, which constitutes an upper bound for the actual value of $\beta_l$ (see SM).
}

\begin{tabular}{c|cc|cc|cc}
\hline
 &   \multicolumn{2}{c|}{$\beta_T$}  &  \multicolumn{2}{c|}{$\beta_l$}  & \multicolumn{2}{c}{$p^C_l$} \\

  &           G & L            &   G &               L     &        G &           L       \\
\hline
$H$	& 	   0.045 &   0.102 &	0.045 &	0.102 &	0.77 &	0.97  
\\
$F$	& 0.008 &  0.018 & 0.008 &	0.018 & 0.23 &	0.44
\\
$W$	& 0.054 &   0.124 &	[0.003, 0.054] & [0.008, 0.124] & [0.11, 0.83]&	[0.23, 0.98]
\\
$B$	& 0.018 &   0.004 &	  5.9e-05 &	  1.5e-05 &	   1.9e-03 &	  {4.9e-04}   \\
$Q$	& 0.047 &   0.012&	  1.1e-06&	  2.6e-07&	   3.5e-05&	  8.7e-06  
\\
\hline
\end{tabular}
\end{table}

\begin{table}[H]
\centering
\caption{\label{infections} 
  \textbf{Distribution of infections in the different locations.}
  Percentage of the total infections that occur in each specific location by the end of the epidemic. 
  The total percentage of infections occurring through frequent contacts ($H, F, W$) and occasional contacts ($B, Q$) are also shown (highlighted) in the table.
  The results correspond to an average of 50 different stochastic realizations of the epidemic, using the same set of parameters for each scenario.
  }
\begin{tabular}{c|cc|cc|}
\hline
Recovery &   \multicolumn{2}{c|}{Exponential}  &  \multicolumn{2}{c|}{Deterministic}  \\
\hline
scenario  &       G &               L     &        G &           L        \\
\hline
$H$	&   33.3\% & 44.0\% &  35.5\% &  47.1\% \\
$F$	&  	8.0\% & 13.9\% &   7.4\% &  12.9\%  \\ 
$W$	&  	18.6\% & 32.1\% &  17.3\% & 30.7\% \\
\rowcolor{lightgray} 
local/ frequent &        59.9\% & 90.0\% &  60.2\% &  90.7\% \\
 $B$	&      8.4\% & 2.1\%   &   8.3\% &  1.9\% \\
$Q$	&    	31.7\% & 7.9\%  &   31.5\% &  7.4\% \\
\rowcolor{lightgray} 
global/ occasional &        40.1\% & 10.0\%  &   39.8\% &  9.3\% \\
\hline
\end{tabular}
\end{table}

\newpage

\end{document}